\documentclass[showpacs,amsmath,amssymb,nofootinbib]{revtex4}

\usepackage{graphicx}
\usepackage{bm}

\newcommand{\br}{{\bf r}}

\begin{document}

\title{
A robust solver for wavefunction-based Density Functional Theory calculations}
\footnote{This manuscript has been authored by UT-Battelle, LLC, under contract DE-AC05-00OR22725 with the US Department of Energy (DOE).
The US government retains and the publisher, by accepting the article for publication, acknowledges that the US government retains a nonexclusive, paid-up, irrevocable, worldwide license to publish or reproduce the published form of this manuscript, or allow others to do so, for US government purposes. DOE will provide public access to these results of federally sponsored research in accordance with the DOE Public Access Plan (http://energy.gov/downloads/doe-public-access-plan).}

\author{J.-L. Fattebert}

\affiliation{Computational Sciences and Engineering Division,
Oak Ridge National Laboratory}

\date{\today}


\begin{abstract}
A new iterative solver is proposed to efficiently calculate
the ground state electronic structure in Density Functional Theory calculations.
This algorithm is particularly useful for simulating
physical systems considered difficult to converge by standard solvers,
in particular metallic systems.
The effectiveness of the proposed algorithm is demonstrated on various applications.
\end{abstract}

\pacs{71.15.-m 71.15.Dx 71.15.Ap 71.15.Pd}

\maketitle


\section{Introduction}

In recent years, growing computational resources have changed the way computer simulations are carried out,
in particular in the domain of Density Functional Theory (DFT) \cite{HK:1},
and its applications to materials science, physics, chemistry, and biology.
Instead of focusing on a single physical system and its physical properties, one can now use high-throughput calculations
to explore many systems, searching for the best candidate for a specific target property, 
or with the goal of designing and optimizing materials properties
\cite{JAIN20112295,Curtarolo2013,Saal2013}.
The search for atomic potentials through Machine Learning (ML) techniques also involves solving the DFT equations numerous
times for various atomic configurations \cite{Behler2016,Deringer2019}.
These configurations are sometimes even random \cite{PhysRevLett.120.156001}.
While innovative work flows have been developed to spawn numerous DFT calculations in an intelligent manner 
(see \cite{MontoyaPersson2017} for instance),
all of these techniques rely on robust solvers for the DFT equations.

The electronic structure of transition metal compounds, open-shell transition metals or heavy elements
can be difficult to converge reliably.
Metallic systems constitute a particular numerical challenge for large-scale DFT calculations, 
due to the strong non-linearity of the Hamiltonian.
This effect can become even more pronounced when dealing with large systems,
or for ''high-Z metals" ({\it i.e.}, those of high atomic number),
as recent advances in computer power allows simulations not only of small unit cells,
but larger systems with hundreds of atoms.
This strong non-linearity can lead to slower convergence for standard iterative solvers,
or even non-convergence and thus limits our simulations capabilities in that field.
The phenomenon know as ``charge sloshing'' is typically worse for large cells \cite{PhysRevB.23.3082}.
To prevent sloshing instabilities the new electronic density can be mixed with a fraction of the
previous density.
This leads users to spend some effort determining, if possible, iterative solver parameters that give reliably convergence for the class of compounds they are interested in.
While convergence failures are often not reported in the literature if they can be fixed ``manually'' this issue is gathering more attention
as we move away from single application/single set of parameters type simulations.  
Techniques to overcome convergence failures during first-principles molecular dynamics have been proposed \cite{WangSong2019},
and ML models have been developed to predict failures \cite{Kulik2019}.

Many research papers have been published in recent years proposing innovative algorithms to solve 
the difficult problem of converging self-consistent iterations in difficult systems.
Most published work however addresses semiconductor systems and systems of much smaller sizes than those we can address today
\cite{PhysRevB.38.12807,PhysRevB.64.121101}.
Here the focus is on direct solvers for wavefunctions based discretizations, that is solvers that attempt at solving directly the non-linear Kohn-Sham (KS) equations 
by computing a set of wavefunctions and their occupations, without
solving surrogate linear eigenvalue problems associated with a frozen Hamiltonian.
Wavefunctions based discretizations include the popular Plane Waves approach or a Finite Difference discretization of the KS equations.
In this category of direct solvers, a very elegant one was proposed by Marzari, Vanderbilt, and Payne
(MVP) \cite{MVP:1}.
It is based on an energy functional that depends on two sets of independent degrees of freedom, the fundamental variables to optimize for:
the electronic wavefunctions and their occupations.
The new solver proposed in this paper is a more robust generalization of the MVP algorithm.
This solver is described in detail in Section \ref{se:KS}.
Some important implementation details are given in Section \ref{se:implementation} and numerical results showing the convergence properties
of the iterative solver are shown in Section \ref{se:results}.

\section{Kohn-Sham solver}
\label{se:KS}

Kohn-Sham theory \cite{KS:1} is a widely used model for first-principles calculations.
It states that the electronic ground state of a physical system can be described by a
system of orthogonal one-particle electronic wave functions
$\phi_j, j=1,\cdots,N$ that minimizes the KS total energy
functional $E_{KS}$.
For a molecular system in a computational domain $\Omega$, the KS
energy functional can be written as (in atomic units)
\begin{eqnarray}
\nonumber E_{KS}[\{\phi_i\}_{i=1}^N,X] &=&
\sum_{i,j=1}^N X_{i,j}
\int_{\Omega}{\phi_i^*(\br)\left(-\frac{1}{2}\nabla^2\right)\phi_j(\br)d\br}
\\ \label{eq:Eks}
&& +\frac{1}{2} \int_{\Omega}\int_{\Omega}{
\frac{\rho_e(\br_1)\rho_e(\br_2)}{|\br_1-\br_2|}d\br_1 d\br_2 }
\\
&+& E_{XC}[\rho_e]
+ \nonumber
\sum_{i,j=1}^N X_{i,j}\int_{\Omega}\phi_i^*(\br)(V_{ext}\phi_j)(\br)d\br.
\end{eqnarray}
The first term represents the kinetic energy of the electrons.
The second one represents the electrostatic energy of interaction between
electrons and is typically computed by solving the corresponding Poisson problem.
$E_{XC}$ models the exchange and correlation between electrons.
Finally, the last term represents the energy of the electrons in the potential $V_{ext}$ generated by
all the atomic cores. 
The electronic density is defined as
\begin{equation}
\label{eq:rho}
\rho_e(\br)=\sum_{i,j=1}^N X_{i,j} \phi_i(\br)\phi_j(\br)
\end{equation}
where $X$ is a single particle density matrix (DM) which describes the occupation of the electronic wave functions
$\phi, i=1,\dots,N$.
To facilitate the later discussion, $X$ is not assumed diagonal here, even though it could be made
diagonal for a particular choice of $\phi_i, i=1,\dots,N$ ---
the eigenfunctions of the Kohn-Sham Hamiltonian --- as it is done most of the time when introducing the KS equations.
$X$ is subject to the constraints $Tr(X)=n_e$ where $n_e$ is the total number of electrons
in the system, and its eigenvalues, the occupations numbers $f_i$ of the electronic eigenfunctions,
are in the interval $[0,1]$.

At finite electronic temperature $T_e$, the functional to minimize is the free energy functional
$A=E_{KS}-T_e S$\cite{Mermin65, MVP:1} which depends on both the wave functions $\{\phi_i\}_{i=1}^N$ and
the occupation numbers $\{f_i\}_{i=1}^N$ --- through $X$ ---
\begin{equation}
\label{eq:Mermin}
A_N[\{\phi_i\}_{i=1}^N,X]=E_{KS}[\{\phi_i\}_{i=1}^N]-T_eS[X]
\end{equation}
where the entropic term is given by
$$
S[X]=\sum_{i=1}^N [-f_i ln(f_i)-(1-f_i)ln(1-f_i)]
$$
and $0\leq f_i \leq 1, i=1,\dots,N$ are the eigenvalues of $X$.
At this functional's minimum, the occupations numbers follow a Fermi-Dirac thermal distribution
\begin{equation}
\label{eq:fT}
f_i=f_T(\epsilon_i-\mu)=\frac{1}{e^{(\epsilon_i-\mu)/kT_e}+1}
\end{equation}
where $\mu$ is the chemical potential whose value is such that the sum of the occupation numbers $f_i$
is equal to the total number of electrons $n_e$.

Because of the non-linearity and the typical problem size, one commonly uses iterative solvers to search for the minimum of the energy functional (\ref{eq:Mermin}).
This generally requires as a key ingredient the steepest descent direction, which can be combined with a preconditioner $K$.
Denoting the set of discretized trial wave functions $\{\phi_i\}_{i=1}^N$ as an $M\times N$ matrix $\Phi$,
these search directions can be written as
$$
-K\nabla_\Phi E_{KS}.
$$
also an $M\times N$ matrix whose matrix elements are the gradient of $E_{KS}$ with respect to each
matrix element of $\Phi$.
We will use the notation $col_j(\Phi)$ to denote the $j^{th}$ column of a matrix $\Phi$,
and $row_i(\Phi)$ to denote the $i^{th}$ row of a matrix $\Phi$.
Here we have $col_j(\Phi)=\phi_j$.

For insulators or semiconductor systems at low temperatures, with a non-vanishing band gap,
numerous solvers are being used in the scientific community,
including various flavors of non-linear conjugate gradient or accelerated gradient methods,
such as the Direct Inversion of the Iterative Subspace (DIIS) \cite{Kresse96}.
Typically, one works with one set of wave functions which are updated iteratively until convergence is achieved.
Probably the most popular approach for dealing with metallic systems consists in solving a series of surrogate problems with linearized operators and occupations numbers set to satisfy (\ref{eq:fT}).
Their solutions are then combined together using some extrapolation scheme until convergence.
Convergence of this process is not guaranteed, is very system dependent, and sometimes problematic for reasons often not well understood.
In practice, it can lead to a lot of parameter tuning, trial, and errors.

Marzari et al. \cite{MVP:1} proposed to solve directly the minimization problem (\ref{eq:Mermin})
by an iterative process where one alternates between wave functions updates --- 
for fixed occupation numbers --- and occupation numbers optimization --- for fixed wave functions.
This was found to be the most robust solver by Woods et al. \cite{Woods_2019}, albeit often a slow one.
One issue with wave function updates for a system with fractional occupation numbers is that the
gradient of the energy functional with respect to an eigenfunction includes the occupation number of that
eigenfunction as a multiplying factor.
This means vanishing corrections for wave functions associated with small occupation numbers and results in slow convergence.
In \cite{MVP:1} Marzari et al. proposed to precondition the gradient in a way that results in calculating gradients as if all the wave functions were equally (fully) occupied.
We observed however that such an approach leads to numerical instabilities, with relatively large updates
of the wave functions near the Fermi level, which leads to significantly reduce step length in the line
minimization procedure in order to keep the iteration stable.
This results in slow convergence as well.

To overcome this problem, we propose to use gradient-based corrections not directly to update the wave functions in an additive fashion,
but instead to use these gradient-based corrections to extend the subspace in which to optimize the single particle density matrix.
That is we propose to improve the MVP algorithm by combining the idea of occupation numbers optimization
with the restarted Davidson algorithm idea of extending the search subspace for wave functions
\cite{DavidsonLiu}.
We find this modification and extension of the algorithm proposed by Marzari et al. to be more robust, precluding the convergence issues observed with the original formulation.

The Davidson algorithm is widely used in the electronic structure community, but typically
to solve an eigenvalue problem with a linear (frozen) Hamiltonian at each step of a self-consistent solver.
Here we propose to apply it to solve for the non-linear Hamiltonian problem, updating the electronic density
and the nonlinear part of the KS potential when optimizing the DM.
Yang et al. \cite{ChaoYang06} proposed a similar approach for systems with a finite band gap at $T_e=$0,
to deal with difficult semiconductor systems,
without the need to optimize the occupations numbers.
They actually propose to use a 3N-dimensional space made not only of the trial vectors and their preconditioned gradients,
but also the trial vectors from the previous iteration, as in the Locally Optimal Block Preconditioned Conjugate
Gradient (LOBPCG) algorithm \cite{knyazev02toward}.
The algorithm presented here generalizes this idea to metallic systems without a band gap.

Let us consider a trial solution $\Phi_k$, an $M\times N$ matrix whose columns are the discretized wave functions $\phi_k, i=1,\dots,N$ at step $k$.
Let us also assume we have an initial guess $X_k$ for the single particle density matrix
associated with that trial solution.
Knowing $\Phi_k$ and $X_k$, one can compute $\rho_{e,k}$, the potential $V(\rho_{e,k})$,
and thus the Hamiltonian operator at step $k$, $H_k$.
Given a preconditioner $K$,
we compute the matrix of preconditioned steepest descent vector projected out of the the subspace spanned by $\Phi_k$ as
$P_k=ortho ((I-\Phi_k\Phi_k^T)K(H_k\Phi_k-\Phi_k (\Phi_k^T H_k\Phi_k)))$,
where $ortho (A)$ denotes a matrix obtained by orthonormalizing the columns of $A$.
Here it means that the matrices $\Phi_k$ and $P_k$ together consistute a set of $2N$ orthonormal vectors.
In the subspace spanned by $\Phi_k$ and $P_k$, one can build the projected
$2N\times 2N$ Hamiltonian matrix
\begin{equation}
\label{eq:Hk2N}
\tilde H_{k,2N}=\left(
          \begin{array}{cc}
            \Phi_k^T H_k\Phi_k & \Phi_k^T H_k P_k \\
            P_k^T H_k\Phi_k & P^T H_k P_k \\
          \end{array}
        \right)
\end{equation}

Then, as proposed by Marzari et al., but in a space twice as large, we compute a search direction for improving $X_k$ based on the solution of
the $2N\times 2N$ symmetric eigenvalue problem
\begin{equation}\label{eq:evp2N}
    \tilde H_{k,2N} Z=Z\Lambda
\end{equation}
where $\Lambda$ is a diagonal matrix with diagonal entries $\epsilon_i, i=1,\dots,2N$ in increasing order, $\epsilon_i\leq\epsilon_{i+1}$.
From the solution of that eigenvalue problem, one then computes a single particle density matrix
\begin{equation}\label{eq:Xk+1}
    \bar X_{2N}=ZF Z^*
\end{equation}
where $F$ is a diagonal matrix with entries $f_i=f_T(\epsilon_i-\mu)$.
$\bar X_{2N}$ is a DM in the space spanned by $\Phi_k$ and $P_k$.

We then define
$$
A_{k,2N}(\beta):=A_{2N}[\{\Phi_k,P_k\},\tilde X_{2N}+\beta(\bar X_{2N}-\tilde X_{2N})]
$$
which is the energy functional (\ref{eq:Mermin}) in a 2N-dimensional space.
Note that in that 2N-dimensional space the $\tilde X_{2N}$ is initialized from $X_k$ as
$$
\tilde X_{2N}=\left(
          \begin{array}{cc}
X_k & 0 \\
0 & 0 \\
\end{array}
        \right).
$$
We do a line search to find the optimal density matrix along the line direction, looking for
\begin{equation}\label{lineSearchX}
    min_{0\leq\beta\leq 1}A_{k,2N}(\beta)
\end{equation}
To do that, we calculate $A_k[\beta=0]$, $A'_k[\beta=0]$ and $A_k[\beta=1]$ and find the minimum of the quadratic polynomial fitting these three values.
$A_k[\beta=0]$ is easy to compute since we already have $\rho_e$ and the potential for the pair $\{\Phi_k,X_k\}$.
To compute $A_k[\beta=1]$, we need to calculate $\rho_e[\{\Phi_k,P_k\},\bar X_{2N}]$, given explicitly by
\begin{equation}
\label{eq:rho2N}
\bar\rho_e(\br)=\sum_{i,j=1}^{2N} (\bar X_{2N})_{i,j} v_i(\br)v_j(\br)
\end{equation}
where
$v_m:=col_m(\Phi_k), m=1,\dots,N$, and $v_m:=col_{m-N}(P_k), m=N+1,\dots,2N$.
From that electronic density, we can evaluate $E_{XC}$ as well as the Coulomb energy.
The other terms in the energy are linear and can simply be computed using the matrix elements associated
with the pairs of wavefunctions and the atomic potentials.
We compute $A'_k[\beta=0]$ partially by finite differences, evaluating $S'_k[\beta=0]=(S_k[\beta=\epsilon]-S_k[\beta=0])/\epsilon$, and then
$A'_k[\beta=0]=Tr(H_{k,2N}(\bar X_{2N}-\tilde X_{2N}))-TS'_k[\beta=0]$.
This way we avoid two problems encountered in the MVP approach:
(i) there is no need to evaluate the derivative of the entropy function, which is numerically problematic since it involves evaluating logarithms close to 0 and 1, and
(ii) this derivative takes fully into account the subspace rotation in the line search direction, which is apparently neglected in MVP.
Note that evaluating the entropy $S_k$ for a specific value of $\beta$ involves computing the eigenvalues
for the corresponding DM.

From the result of that line minimization problem, $\beta^*$, one obtains a new trial DM in the 2N-dimensional space
given by
\begin{equation}
\label{eq:betastar}
\tilde X_{2N}\gets\tilde X_{2N}+\beta^*(\bar X_{2N}-\tilde X_{2N}).
\end{equation}
Using the updated $\tilde X_{2N}$, a new electronic density is computed 
\begin{equation}
\label{eq:rho2N}
\rho_e(\br)=\sum_{i,j=1}^{2N} (\tilde X_{2N})_{i,j} v_i(\br)v_j(\br)
\end{equation}
where
$v_m:=col_m(\Phi_k), m=1,\dots,N$, and $v_m:=col_{m-N}(P_k), m=N+1,\dots,2N$.

The above process of optimizing the DM within a 2N-dimensional space can be repeated several times,
from defining a new DM target using Eqs. (\ref{eq:evp2N}) and (\ref{eq:Xk+1}),
to doing a new line minimization between the latest DM and this new target, to updating DM.
In practice, it is not necessary to converge that process and a fixed number of line minimizations (2 or 3) is sufficient.
Once this iterative process is complete, we can extract a subset of N vectors out of the 2N spanning the subspace
in which we optimized $X$.
To do that, the last $\tilde X_{2N}$ is diagonalized, solving the symmetric eigenvalue problem
\begin{equation}\label{eq:evpX2N}
    \tilde X_{2N} V=V\Lambda
\end{equation}
where $\Lambda$ is a diagonal matrix with real diagonal entries $\lambda_i, i=1,\dots,2N$ in increasing order, $\lambda_i\leq\lambda_{i+1}$,
and keep the N vectors corresponding to the highest eigenvalues $\lambda_i, i=N+1,\dots,2N$, that is
\begin{equation}
\Phi_{k+1}=\left(\Phi_k,P_k\right)V_2
\end{equation}
where $V$ has been split into two $2N\times N$ matrices, $V=\left(V_1,V_2\right)$.
The $N$ highest eigenvalues should all be between 0 and 1, corresponding to the fully and partially occupied states of $X_{k,2N}$.
The discarded eigenvalues should all be 0, corresponding to fully empty states:
if they are not, it would mean more empty states should be included in the calculations to fully cover all the partially occupied states.
Note that the diagonalization of the density matrix can lead to several degenerate eigenvalues --- around 0 ---
and thus 
lead to eigenvectors being very sensitive to numerical noise or the specific algorithm used for diagonalization.
While this does not affect the final result, it does affect intermediate results before convergence
for this solver.

Once we are back to a set $\Phi$ of $N$ vectors describing the electronic structure of the system, we can start again and evaluate
a gradient-based search direction,
and generate a new $2N$-dimensional space in which to optimize the DM again.
We will refer later to this part as the ``outer'' iteration, while the DM optimization inside that $2N$-dimensional space is referred to as the ``inner'' iteration.

While we do not have convergence proof for the algorithm described above,
by construction it should lead to non-increasing energies since every update is based on an energy minimization.
If $A'_k[\beta=0]<0$, one should have $\beta^*>0$ and the energy should be decreasing.
However, while this property has been proven for systems with band gaps at $T_e$=0 \cite{cances2001},
to the best of the author’s knowledge it has not been demonstrated for $T_e>$0.

\section{Implementation}
\label{se:implementation}

The solver described in the previous section was implemented in the MGmol code,
an open source finite difference DFT code \cite{mgmol}.
Because the solver relies on an energy minimization (\ref{lineSearchX}), it is important for the discretized KS potential
to be consistent with the discretized exchange-correlation energy.
This comes out naturally for real-space discretization of the Local-density approximations (LDA)
exchange-correlation with point-wise functional derivatives.
For the Perdew–Burke-Ernzerhof (PBE) \cite{PBE:1} functional,
the exchange-correlation potential proposed by White and Bird \cite{WhiteBird1994} is used.
For a discrete set of grid points and a finite difference evaluation of the gradients of the electronic density,
it is consistent with the discretized form of the exchange-correlation energy.
Besides that, there is nothing specific to a real-space discretization in the solver and convergence results are expected to be the same using
a different discretization, such as the popular Plane Waves approach.

Given that the algorithm described above involves calculations in a 2N-dimensional space that can become
quite expensive compared to simple updates in an N-dimensional space, it can be important to pay attention
to some implementation details that may slow down the solver considerably in an unoptimized implementation.
For instance, evaluating the electronic density $\rho_e$ according to
Eqs. (\ref{eq:rho}) or (\ref{eq:rho2N}) can be computationally expensive given the double sum over the electronic states,
in particular when the sum goes up to $2N$, when evaluating $A_k$ at $\beta=0$, and $\beta=1$.
One notices however that the number of floating point operations necessary to carry out that sum is $M N^2$,
thus the same as computing the overlap matrix $S=\Phi^T \Phi$ for example.
In practice however, while optimized libraries exists on most computing platforms to carry out matrix-matrix multiplications
very efficiently, no such functionality exists for the expression (\ref{eq:rho}) since it is not a standard linear algebra operation.

To take advantage of vendor optimized Basic Linear Algebra Subprograms (BLAS) libraries \cite{blas3}, let us rewrite (\ref{eq:rho})  in a block form.
Let $\boldsymbol\rho_e$ denote the vector made of the values of $\rho$ at all the mesh points $\bf r$.
It can be expressed as
\begin{equation}
\label{eq:rhov}
{\boldsymbol\rho_e}=diag(\Phi X \Phi^T)
\end{equation}
where $diag(A)$ denotes a column vector consisting of the diagonal entries of matrix $A$.
Its components $i$ at ${\bf r}_i$ is then given by
\begin{equation}
\label{eq:rhoi}
{(\boldsymbol\rho_e)}_i=
 \left(row_i(\Phi X)\right)^T \cdot row_i(\Phi).
\end{equation}
To be computationally efficient, the idea is to first evaluate the matrix product $\Phi X$, which requires $M N^2$ floating point iterations,
followed by the $M$ dot products of row vectors in Eq. (\ref{eq:rhoi}) which require each $N$ floating point iterations each, for a total of $MN$ operations.
The first operation is the expensive one --- $O(N^3)$ for $M\sim N$ --- and can be carried out with an optimized matrix-matrix product implementation (BLAS3).
The second one requires much fewer operations --- $O(N^2)$ for $M\sim N$ --- and can be computed with $M$ dot products (BLAS1).

Another place where computational efficiency can be enhanced is by using Kleinman-Bylander (KB) separable pseudopotentials \cite{KB82}.
Indeed, in the evaluation of the matrix elements of $\tilde H_{k,2N}$ in Eq. (\ref{eq:Hk2N}), the linear part corresponding to the pseudopotentials
can then be computed using a set of pre-computed dot products $<\beta_{lm},\phi_i>$ between all the KB projectors from all of the atoms and all of the 2N
wavefunctions, which can considerably simplify the whole calculation.

The solver described in the previous section was also implemented for systems with spin.
In that case, one has a DM for each spin.
The line minimization however should be done for the overall energy of the total system, resulting in a single
$\beta^*$ in (\ref{eq:betastar}).
An application with spin is shown in Section \ref{se:results} to illustrate that case.
This strategy should be used for calculations with k-points too, but has not been implemented and will not
be shown here.

\section{Numerical Results}
\label{se:results}

A series of test problems was setup to evaluate the iterative solver proposed above.
The focus is on the algorithm performance and no particular attention was made to pick the optimal DFT functional
for each system from a physics point of view.
The list of test problems and their parameters are described in Tab. \ref{tab:systems}.
These includes finite systems, bulk, and surfaces.
The geometry for the spin-crossover complex Fe(salen)(NO) was taken from \cite{FeSalenNO}.
The geometry for the $\mathrm{Au_{13}}$ cluster is from \cite{Kinaci2016} ($\mathrm{Au_{13}(1)}$),

The Optimized norm-conserving Vanderbilt pseudopotentials (ONCVPSP) as proposed by Hamann \cite{Hamann2013} were used
for all the numerical experiments shown here.
For the simulations using the PBE exchange and correlation functional, the ONCVPSP potentials as parameterized by 
Schlipf and Gygi \cite{SCHLIPF201536} were used.
The number of valence electrons/atom for these pseudopotentials are 1 for H, 4 for C, 5 for N, 6 for O, 16 for Fe, 19 for Au, 19 for Cu, and 11 for La.
Note that for all the simulations presented here except for the spin-crossover complex Fe(salen)(NO),
that is the systems without spin,
the maximum number of electrons/wavefunction is set to two to take advantage of degeneracy and reduce computational cost.

\begin{table}
\begin{tabular}{ |c|c|c|c|c|c|c|c|c| } 
 \hline
system & type & computational domain & XC & spin & \# w.f. & mesh spacing 
& \# inner iterations & $\mathrm{T_e}$(K) \\ 
 \hline
$\mathrm{F}$ & single atom & 21. x 21. x 21. & PBE & no & 5 & 0.33 & 2 & 300
\\
  $\mathrm{Au_{13}}$ & molecule & 56.7 x 56.7 x 28.35 & LDA & no & 174 & 0.29 & 3 & 300
\\ 
$\mathrm{Au_{32}}$ & surface & 15.4 x 15.4 x 30.8 & LDA & no & 404 & 0.24 & 3 & 600
\\
 Fe(salen)(NO) & molecule & 25.6 x 25.6 x 38.4 & PBE & S=1 & 69 & 0.2 & 2 & 300 
\\ 
$\mathrm{Cu{32}}$ & surface & 13.6 x13.6 x 27.2 & LDA & no & 368 & 0.28 & 3 & 300
\\
$\mathrm{Cu_{108}}$ & bulk & 20.4 x 20.4 x 20.4 & LDA & no & 1242 & 0.28 & 3  & 300
\\
$\mathrm{La_{56}}$ & bulk & 28.50 x 24.68 x 22.98 & LDA & no & 358 & 0.2 & 3 & 300 
\\
 \hline
\end{tabular}
\caption{List of physical systems used to evaluate convergence rates.
The unreconstructed ``surfaces'' were generated by simply extending the computational domain for a cubic cell made of 32 atoms by a factor 2 in one direction.}
\label{tab:systems}
\end{table}

\begin{figure}
\includegraphics[scale=0.4]{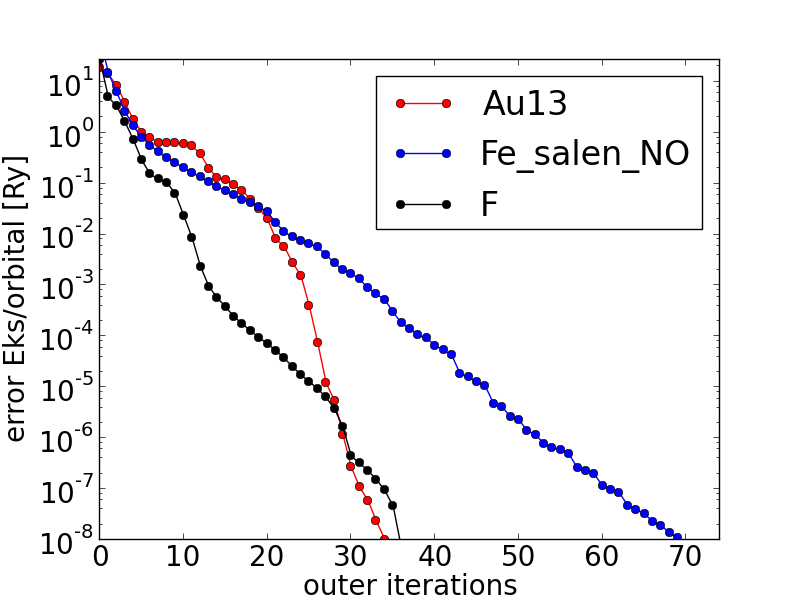}
\caption{Convergence as a function of number of outer iterations for three finite systems.
The reference energies are the values obtained after a tight convergence of the iterative solver.}
\label{fig:finite}
\end{figure}

\begin{figure}
\includegraphics[scale=0.4]{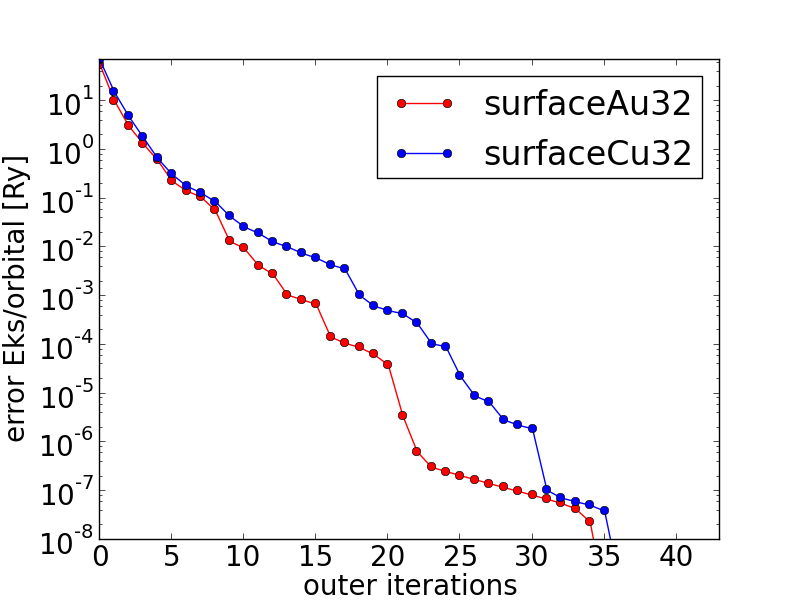}
\caption{Convergence as a function of number of outer iterations for two surfaces (Gold and Copper).
The reference energies are the values obtained after a tight convergence of the iterative solver.}
\label{fig:surface}
\end{figure}

\begin{figure}
\includegraphics[scale=0.4]{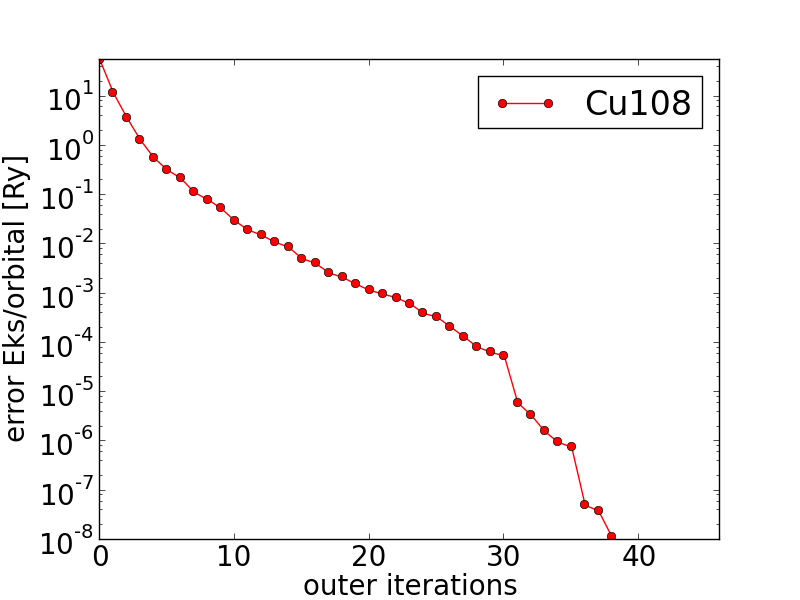}
\caption{Convergence as a function of number of outer iterations for bulk Copper (108 atoms cell).
The reference energies is the value obtained after a tight convergence of the iterative solver.}
\label{fig:cu108}
\end{figure}

\begin{figure}
\includegraphics[scale=0.4]{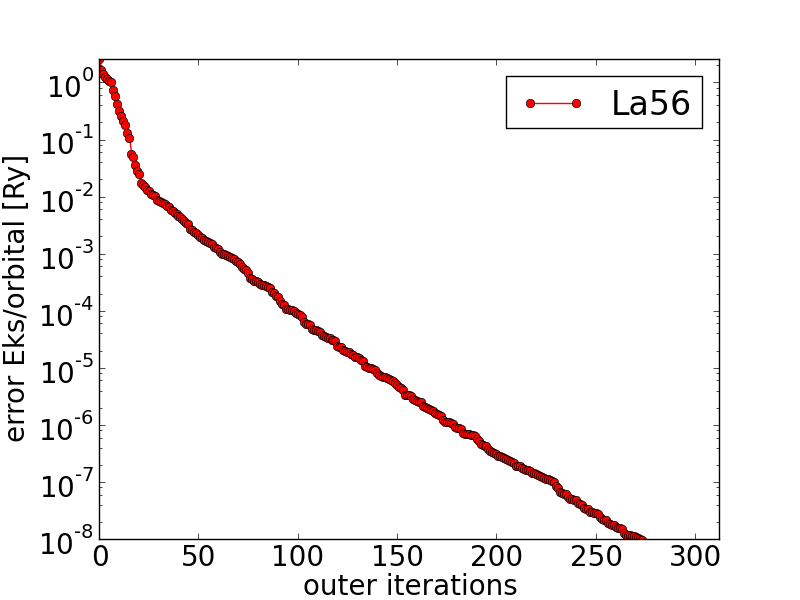}
\caption{Convergence as a function of number of outer iterations for bulk Lanthanum (56 atoms cell).
The reference energy is the value obtained after a tight convergence of the iterative solver.}
\label{fig:la56}
\end{figure}

Fig. \ref{fig:finite} shows convergence of the iterative solver for three finite systems: a gold cluster,
an open shell transition metal complex, and a single F atom.
Fig. \ref{fig:surface} shows convergence for two metallic surfaces: gold and copper.
Fig. \ref{fig:cu108} shows convergence for a really large bulk system with 108 copper atoms and
1242 wavefunctions.
Finally Fig. \ref{fig:la56} shows convergence for a bulk system made of 56 Lanthanum atoms.
It can be observed that most systems converge to a tolerance of $10^{-8}$ Ry/wavefunction within less than 70 outer iterations.
The exception is $\mathrm{La_{56}}$ which takes more than 200 outer iterations to converge to that same tolerance.
As expected from the algorithm properties, convergence is monotonic for all systems.

\section{Concluding remarks}

A new algorithm to solve for the Density Functional Theory equations was presented
for wavefunction-based discretizations.
Its key property is robustness and a guaranteed non-increase of the energy of the system from iteration to iteration.
Convergence is observed to be fast as well, typically requiring less than 100 wavefunction updates.
The computational cost to pay for that is to work in a 2N-dimensional space that include the trial wavefunctions and their correction directions,
but this is similar to what the popular restarted Davidson algorithm requires too, albeit in a non-linear context.

A possible extension of the algorithm could be considered by working in a 3N-dimensional space instead of 2N, similar to
the LOBPCG algorithm \cite{knyazev02toward}, including not only the trial wavefunctions and the correction directions,
but in addition the trial wavefunctions from the previous step.
This would likely improve convergence given the larger search space for the DM optimization.
While time-to-solution is clearly implementation dependent, it would be interesting to see if the extra cost
of a 3N-dimensional space would be compensated for by a substantially smaller number of iterations.

Another interesting study would be to evaluate that algorithm for atomistic systems modeled using
multiple k-points.
While the author does not anticipate any issues in that case, that investigation remains to be done.

\section*{Acknowledgements}

Research sponsored by the Laboratory Directed Research and Development Program of Oak Ridge National Laboratory (ORNL),
managed by UT-Battelle, LLC for the U. S. Department of Energy under Contract No. De-AC05-00OR22725.
This research used resources of the Compute and Data Environment for Science (CADES) at the Oak Ridge National Laboratory,
which is supported by the Office of Science of the U.S. Department of Energy under Contract No. DE-AC05-00OR22725.


\bibliography{biblio}

\end{document}